\documentclass{revtex4}
\usepackage{graphicx}
\begin{document}
\title{The influence of self-citation corrections on Egghe's $g$ index}
\author{Michael Schreiber
 }
 \affiliation{
 Institut f\"ur Physik, Technische Universit\"at Chemnitz, 09107
  Chemnitz, Germany }

\setlength{\tabcolsep}{3.0mm}

\begin{abstract}

  The $g$ index was introduced by Leo Egghe as an improvement
  of Hirsch's index $h$ for measuring the overall citation record
  of a set of articles. It better takes into account the
  highly skewed frequency distribution of citations than the
  $h$ index. I propose to sharpen this $g$ index by excluding the
  self-citations. I have worked out nine practical cases in physics
  and compare the $h$ and $g$ values with and without
  self-citations. As expected, the $g$ index characterizes the
  data set better than the $h$ index. The influence of the
  self-citations appears to be more significant for the $g$ index than
  for the $h$ index.

\end{abstract}

\maketitle

\section{Introduction}
In 2005 the physicist Hirsch \cite{Hirsch} introduced the $h$
index as an easily determinable estimate of the impact of a
scientist's cumulative research contribution. The index $h$ is
defined as the highest number of papers of a scientist that
received $h$ or more citations. Thus it incorporates both
publication quantity and citation quality into a single number,
which might be taken as an estimate of the visibility or impact of
a scientist's research and might reflect its importance and
significance. It has generally been well received in the physics
community \cite{Pop,Leh,Bat} and it is discussed as a convenient
measure in evaluations.\\ \indent The $h$ index is robust in the
sense that it is insensitive to the number of uncited or lowly
cited papers. This is certainly an advantage compared to other
indicators like the average number of citations per paper or the
total number of papers. On the other hand, the $h$ index is also
robust in the sense that it is not sensitive to one or several
outstandingly highly cited papers, because once a paper has
reached the $h$-defining set, it is no more relevant whether or
not it is further cited. This aspect has been considered as a
major drawback of the $h$ index \cite{Eg5}. One way to take into
account the performance of the most cited papers would be to
determine the average number of citations per ``meaningful paper``
\cite{Pod}, but then it is ambiguous how to define the threshold.
One possibility would be to use the value $h$ \cite{Pod}, the
result has been labeled $a$ index \cite{Bih,Meho}. Thus the
performance could be characterized by two values, the $h$ index
and the $a$ index. A more elegant way to incorporate the evolution
of citation counts of highly cited articles into a single number
has been proposed by Egghe \cite{Eg2,Eg4}, defining the $g$ index
as the highest number of papers that received on average $g$ or
more citations. In other words, it is the highest number of
articles that together received $g^2$ or more citations. By this
definition the usually strongly skewed frequency distribution of
the citations increases the score: the higher the number of
citations in the top range, the higher the $g$ index.\\ \indent
The relation between $h$ and $g$ has been investigated for some
simple models \cite{Ros}, it has also been applied to 14 Price
medalists \cite{Eg5}. In that paper the author suggested that it
would also be interesting to work out more practical cases (in
other fields) of $h$ and $g$ index comparisons. It is one purpose
of the present article to present nine such practical cases in
physics. Previous investigations of the Hirsch index have
determined outstanding values for prominent and highly cited
scientists. While these are certainly interesting, they are surely
not representative cases. Only one investigation \cite{Roe} of the
Hirsch index analyzed the data of more average scientists, namely
average faculty members. I have chosen the same strategy,
analyzing my own publications and those of eight colleagues as
described below. This case study shows that the $g$ index better
measures the citation records of the authors than the $h$ index.
\\
\indent The second purpose of the present work is to investigate
the influence of self-citations on the $g$ index. Self-citations
do not reflect the impact of a publication and should therefore
not be included in a citation analysis when this is intended to
give an assessment of the scientific achievement of a scientist
and his visibility \cite{Aks}. Earlier studies \cite{Sny} have
shown that in physical sciences 15 \% of all citations were
self-citations. In more recent studies \cite{Aks,Gl4} in physics
about 25 \% of all citations where identified as self-citations.
However, it is not obvious, how strongly these self-citations
influence the $g$ index. I have recently \cite{MS} shown that
self-citations significantly reduce the $h$ index in contrast to
Hirsch's expectations \cite{Hirsch}. For a group of 7 scientists
in ecology and evolution \cite{Kel} the exclusion of
self-citations reduced the $h$ index on average by 12.3 \%. On the
other hand, the average reduction was only 6.6 \% for the $h$
indices of 31 influential scientists in information science
\cite{Cro}, with an absolute decrease between zero and three only.
In contrast, my investigation \cite{MS} showed a decrease between
two and six, with an average of 3.6 or 23.5 \%, when
self-citations were excluded. This is a rather significant effect.
I will show below that the effect is similarly important for the
determination of the $g$ index, yielding a decrease of the $g$
index between one and twelve, with an average of 5.0 or 21.7 \%.
It also turned out that the $g$ index after exclusion of
self-citations allows for a better distinction between the impact
of the publications of the different authors. Thus it is superior
to the original $g$ index, which in turn is superior to the $h$
index.
\\

\section{Data base}
The subsequent analysis is based on data compiled in January and
February 2007 from the Science Citations Index, provided by
Thomson ISI in the Web of Science (WoS), taking great care that
homographs do not distort the results \cite{MS}. Out of
self-interest and due to the fact that it is relatively easy to
check for homographs and other inaccuracies in one's own
publications and citations I first performed a self-experiment and
investigated my own citation records (data set A). Then I analyzed
the publications of a somewhat older colleague who is working in a
more topical field in a mainstream area (data set B). In contrast,
I studied the records of a somewhat younger colleague, working in
a less attractive field, who has published fewer papers (data set
C) and I was surprised that in this case the $g$ index was not
significantly larger than the $h$ index, see below. This induced
me to look at two other data sets (D and E) with an even smaller
total numbers of publications. In these cases the ratio $g/h$
turned out to be unexpectedly large, as also shown below. Finally,
addressing the concerns of a referee, I have considered four more
cases, where F was chosen because it appeared to be similar to D
and E, and G, H, I comparable to C.
\\ \indent Small deviations of the values for data sets A and B in comparison to
the previous study \cite{MS} are due to the fact that the data
have been updated for the present investigation.

\section{Hirsch's index $h$ and Egghe's index $g$}
After ordering the publication list according to the number of
citations $c(r)$, where $r$ is the rank that the ordering has
attributed to the
article,\renewcommand{\thefootnote}{\fnsymbol{footnote}}\footnotemark[4]\footnotetext[4]{The
notation $g(r)$ is commonly used for the general rank-frequency
function, but in order to avoid confusion with the value $g$ for
the $g$ index I rather use the notation $c(r)$ for the citation
frequency of the article with rank
$r$.}\renewcommand{\thefootnote}{\arabic{footnote}} Hirsch's index
$h$ can be easily determined from

\begin{equation} \label{equ:H}
h \le c(h) < h+1,\end{equation} or, equivalently, from
\begin{equation} \label{equ:H1}
h = \max_r \big(r \le c(r)\big), \end{equation} which reflects the
verbal definition above, namely that $h$ is given by the highest
number of papers which received $h$ or more citations. To
determine the $g$ index, one has to calculate the sum $s(r)$ over
the number of citations up to the rank $r$:

\begin{equation} \label{equ:H2}s(r) = \sum\limits^r_{r'=1}
c(r')\end{equation} Then the $g$ index is also easy to determine
from

\begin{equation}
\label{equ:H3}g^2 \le s(g) < (g+1)^2\end{equation} or,
equivalently, from
\begin{equation} \label{equ:H4}g = \max_r \big(r^2 \le s(r)\big)=\max_r
\big(r \le \bar{c}(r)\big)\end{equation} where $\bar{c}(r)$
denotes the average number of citations up to rank $r$, i.e.,
\begin{equation} \label{equ:H7}\bar{c}(r) =
\frac{s(r)}{r}.\end{equation} The definition (5) again reflects
the verbal definition above, namely that the $g$ index is defined
as the highest number of papers which received on average $g$ or
more citations.

\section{Results of the first analysis: $h$, $a$, and $g$ indices}
In table 1 some characteristics for the nine data sets are
compiled. Besides the total number of papers $n$ also the number
of papers with at least one citation, $n_1= \max r(c \ge 1)$, is
given, because in the general framework of information production
processes where the articles are the sources and the citations are
the investigated items, only sources with at least one item are
usually considered. In citation analysis, on the other hand, of
course also uncited articles belong to the data set. But it is
certainly interesting to observe how many articles are not cited
at all. In the cases which are investigated here, between 14 \%
and 31 \% of the papers received no citation. As well it is
interesting to see the highest number of citations for every
author, $c(r=1)$. Here the high value $c^D(1)=204$ for data set D
(The superscript is used to distinguish the data sets.) is
conspicuous in comparison to the relatively small total number of
publications. On the other hand, the value $c^C(1)= 24$ appears to
be quite small for a scientist with an overall production of 86
papers.\\
\begin{table}[t]
\caption{Characteristics for the nine data sets analyzed in the
present investigation. The first column labels the data sets, the
following columns give the total number of publications, the
number of publications which received at least one citation, the
highest citation count, the Hirsch index, the total number of
citations of all papers up to the Hirsch index, the average number
of citations for all papers up to the Hirsch index, the geometric
mean of the Hirsch index and the $a$ index, the $g$ index, the
citation count of the first paper that is not included in the
$g$-defining set, and the ratio of $g$ index and $h$ index.\\}
\label{tab:1}
\renewcommand{\arraystretch}{1.1}
\begin{tabular}{crrrrrrrrrr}
\hline
  \vspace{0cm}
     data set & $n$ & $n_1$ & $c(1)$ & $h$ & $s(h)$ & $a=\bar{c}(h)$ & $\sqrt{s(h)}$ & $g$ & $c(g+1)$ &
     $g/h$\\
     \hline
 A & 270 & 214 & 182 & 27 & 1691 &62.6 & 41.1 & 45 & 18 & 1.67\\
 B & 290 & 250 & 457 & 39 & 3661 &93.9 & 60.5 & 67 & 22 & 1.72\\
 C &  86 &  59 &  24 & 13 &  222 &17.1 & 14.9 & 15 & 10 & 1.15\\
 D &  35 &  29 & 204 &  8 &  281 &35.1 & 16.8 & 18 &  3 & 2.25\\
 E &  15 &  12 &  25 &  5 &   85 &17.0 &  9.2 & 10 &  2 & 2.00\\
 F &  25 &  19 &  19 &  7 &   77 &11.0 &  8.8 & 9  &  5 & 1.50\\
 G &  88 &  67 &  64 & 14 & 428 & 30.6 & 20.7 & 22 & 8  & 1.57\\
 H &  72 &  61 &  55 & 14 & 388 & 27.7 & 19.7 & 22 & 11 & 1.57\\
 I &  79 &  56 &  55 & 14 & 388 & 27.7 & 19.7 & 21 & 10 & 1.50\\
  \hline
  \end{tabular}
\end{table}
The first analysis of the data yielded the Hirsch index as given
in table 1. It is apparent from the strongly differing values of
$h$ that the present study describes rather different cases as
mentioned above.
\\ \indent In the present context, the summed number of citations up to
the value of the Hirsch index, $s(h)$, is significant, because it
yields the $a$ index, given by the average $a=\bar{c}(h)$ as shown
in table 1. As expected, the $a$ index is significantly larger
than the $h$ index, indicating already that one or several papers
have a quite high citation count which cannot be appropriately
appreciated by the Hirsch index. This is of course reflected in
the determination of the $g$ index for the nine data sets (see
table 1) which provides the expected \cite{Eg2} larger variance
compared to the Hirsch index $h$. The geometric mean between the
$h$ index and the $a$ index, which is given by $\sqrt{s(h)}$,
yields already a reasonable estimate for the $g$ index. This means
that the $h$-defining set is strongly dominating the
rank-frequency distribution, the citation counts of the further
publications have only a minor influence. This is due to the fact
that they are already relatively small and therefore cannot
contribute very much to an enhancement of the $g$ index. In the
extreme situation that $c(r)=0$ for all $r>h$ one would get
$g=\sqrt{s(h)}$.\\ \indent Of course, the number of citations at
the critical rank $r=g$ is usually significantly lower than the
index $g$ and also much lower than the value $h$. The numbers
$c(g+1)$ in table 1 show that although for the $g$ index a
significantly longer part of the tail of the citation distribution
has to be taken into account, the tail is not nearly exhausted in
most cases, except data sets D and E. Small values of $c(g+1)$ are
consistent with the above discussion that the tail is long, but
weak and therefore does not contribute much to the enhancement of
the $g$ index. To be specific, the value $c(g+1)$ must be compared
to $2g+1$ which is the number of additional citations which are
needed to increase the index from $g$ to $g+1$.
\\ \indent
The ratio $g/h$ has already been suggested as an interesting
measure \cite{Eg5}, because this relative increase indicates the
existence of very highly cited articles in the data set, where
very high is meant in relation to the Hirsch index. In this
respect data set D is outstanding with $g^D/h^D=2.25$ due to
$c^D(1)=204$ although data set B contains the article with by far
the highest total citation count $c^B(1)=457$. Also outstanding is
data set C, with the smallest ratio $g/h$ which means that all
manuscripts up to the rank $r=h^C$ have citation counts of the
same magnitude. This was already reflected in the $a^C$ index,
which is not significantly higher than the $h^C$ index. Analogous
though less conspicious in this respect is case F. As intended
with the selection of the data sets, the cases G, H, and I appear
to be quite similar.

\section{Self-citations and the sharpened index $g_s$}
In any measure of scientific achievement, self-citations should
not be included, because they do not reflect the visibility or
impact of the scientist's research. Of course, some self-citations
are completely legitimate, for example when they are really needed
in a manuscript in order to avoid repetition of previously
published theoretical models, experimental setups, results, and/or
conclusions. But in some cases the number of self-citations is
relatively high, which might be for the simple reason that authors
find it easier to refer to their own papers when a citation is
required in a given context for a certain argument, because one
knows one's own previous manuscripts best. However, sometimes one
can get the impression that people cite their own papers only,
because nobody else does. It is certainly tempting to enhance
one's citation count when more and more assessments are based on
such a quantity. The Hirsch index is particularly vulnerable to
such practice, because it is a single number, which can be
relatively easily enhanced by specifically citing those papers for
which the citation count is close to but below the critical value
$h$. The $g$ index is also just a single number, but it is less
vulnerable being essentially an integral measure so that a small
change of the citation count of a single manuscript close to the
critical value $g$ has in most cases no effect on the index.
Nevertheless, the integrated number of self-citations does have a
significant effect on the index as
will be shown below.\\
\begin{figure}
\includegraphics[height=5.5cm]{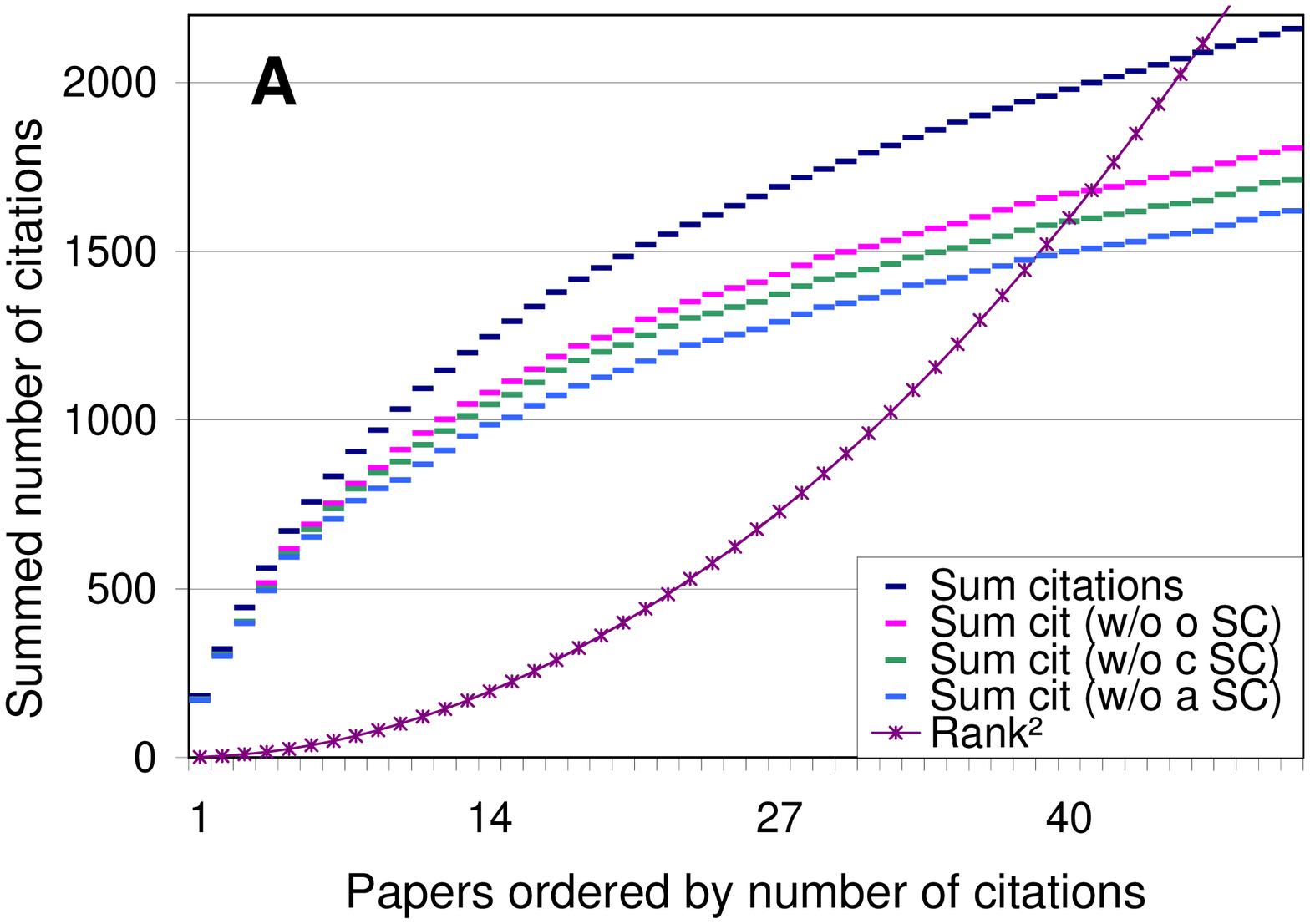}\hspace{2em}\includegraphics[height=5.5cm]{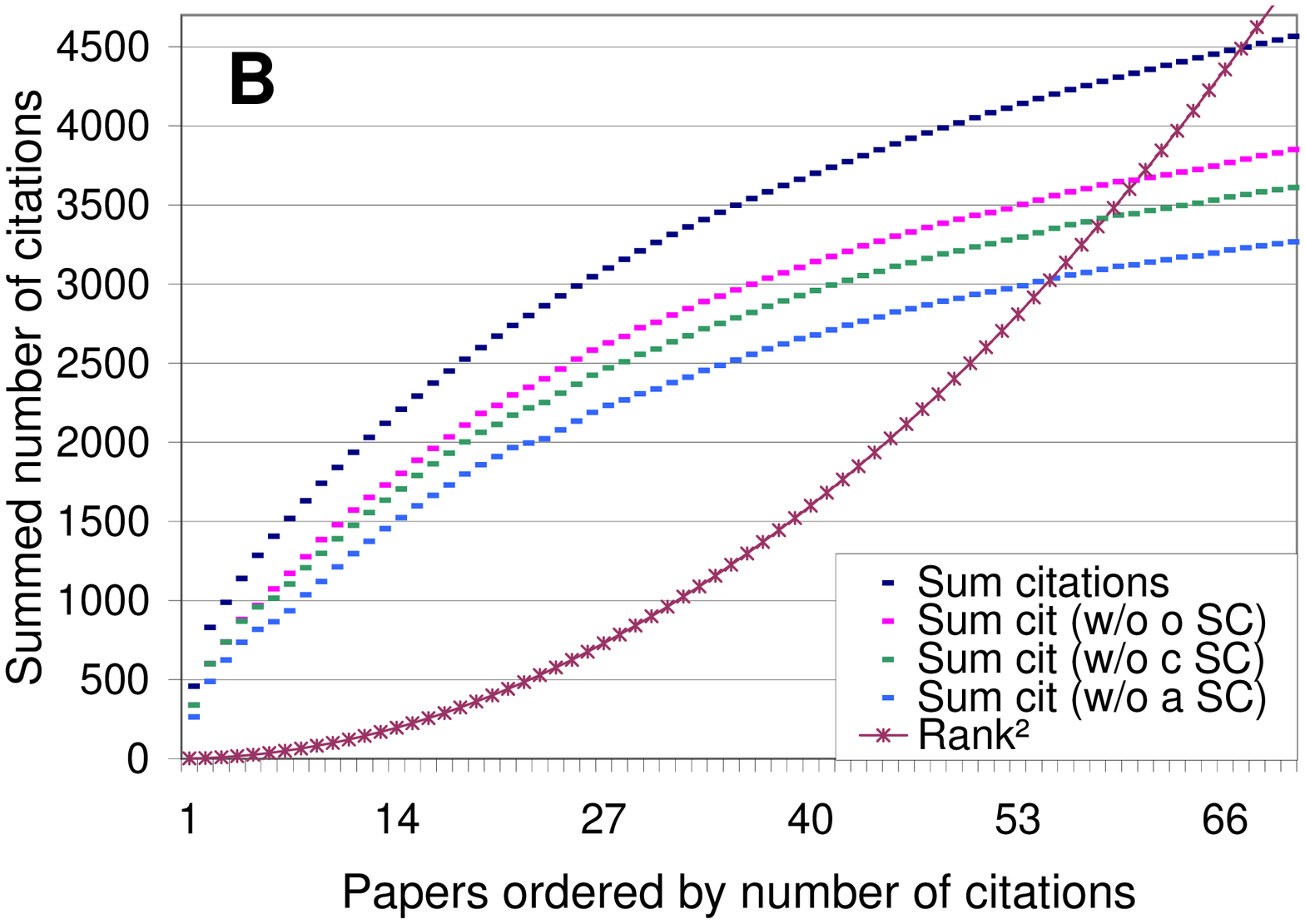}
\\
\vspace{0.5cm}
\includegraphics[height=5.0cm]{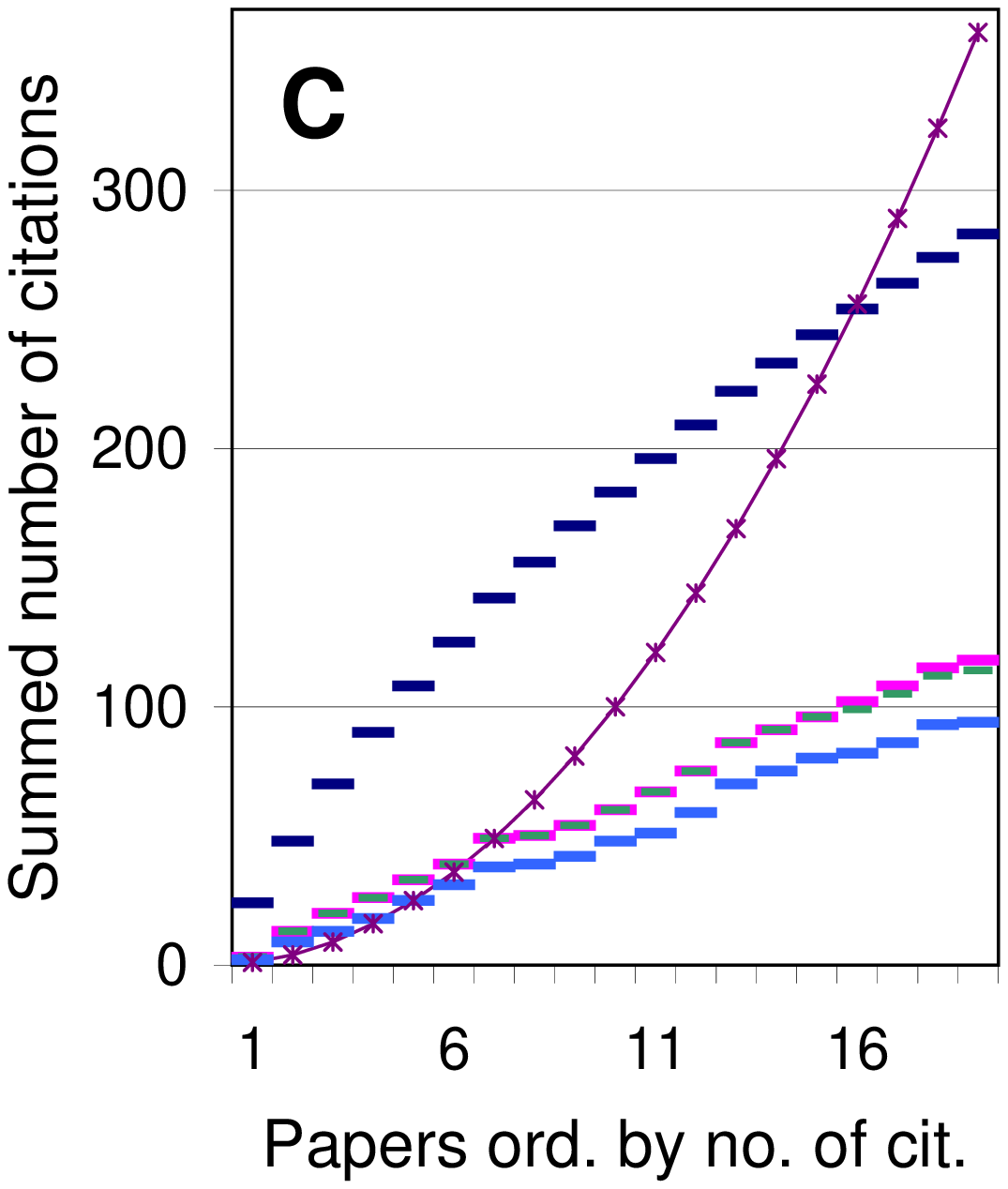}\hspace{2em}\includegraphics[height=5.0cm]{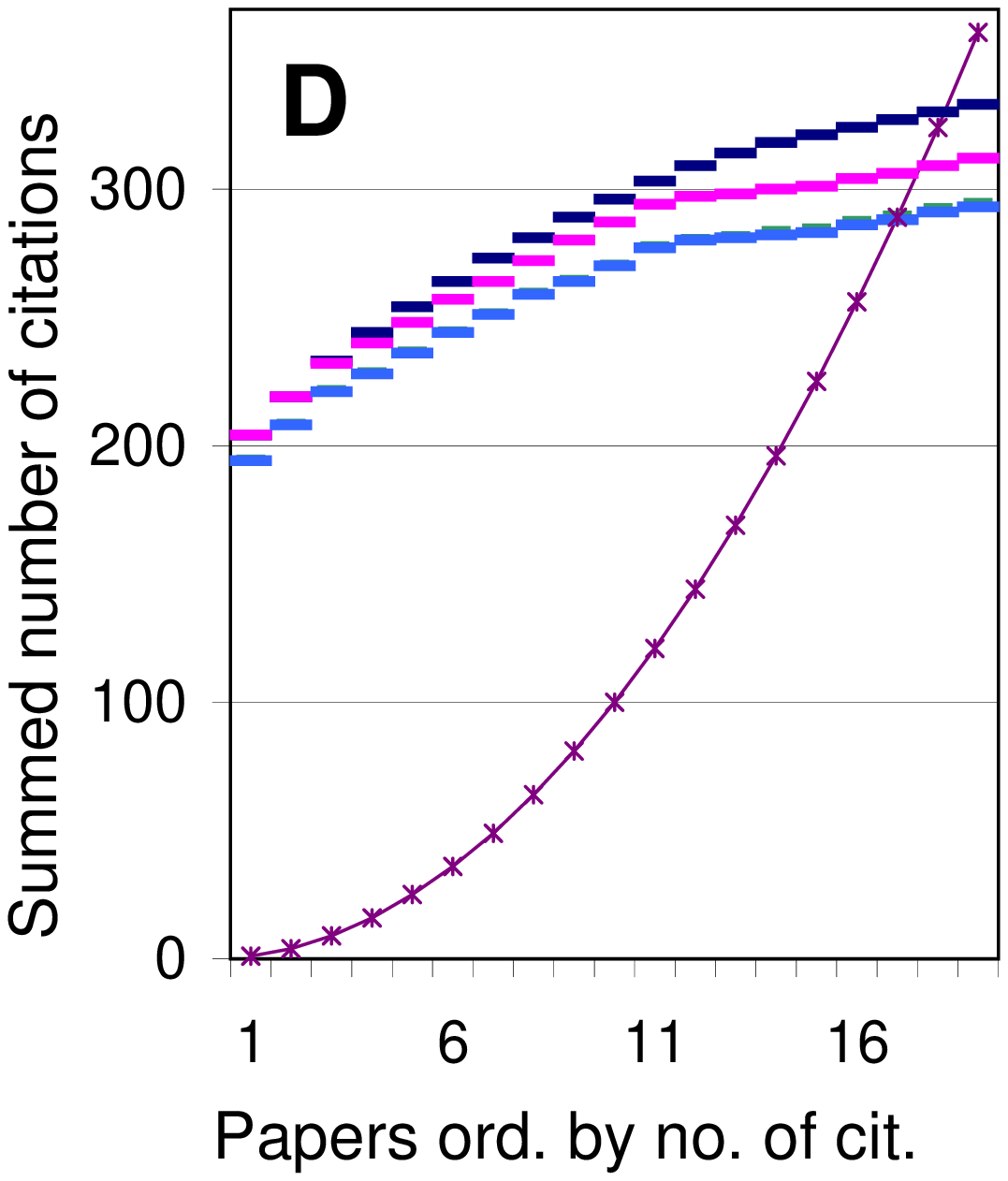}\hspace{2em}\includegraphics[height=5.0cm]{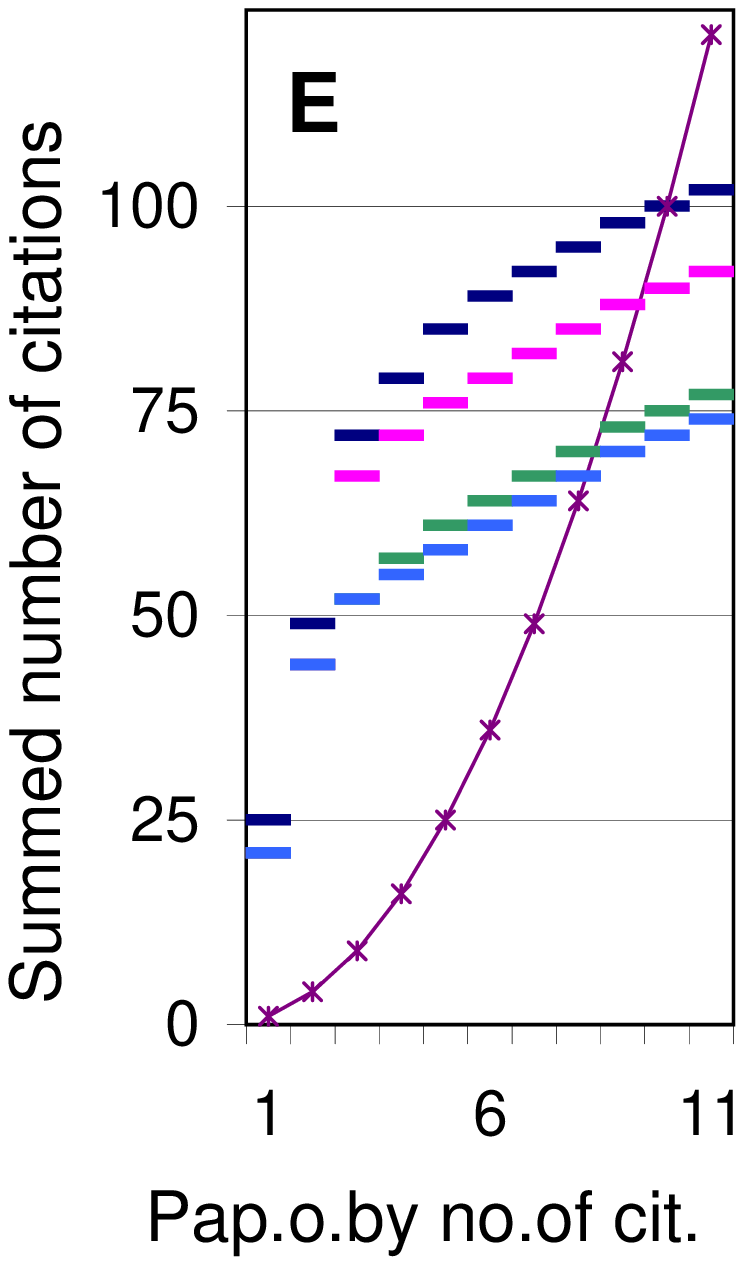}
\hspace{2em}\includegraphics[height=5.0cm]{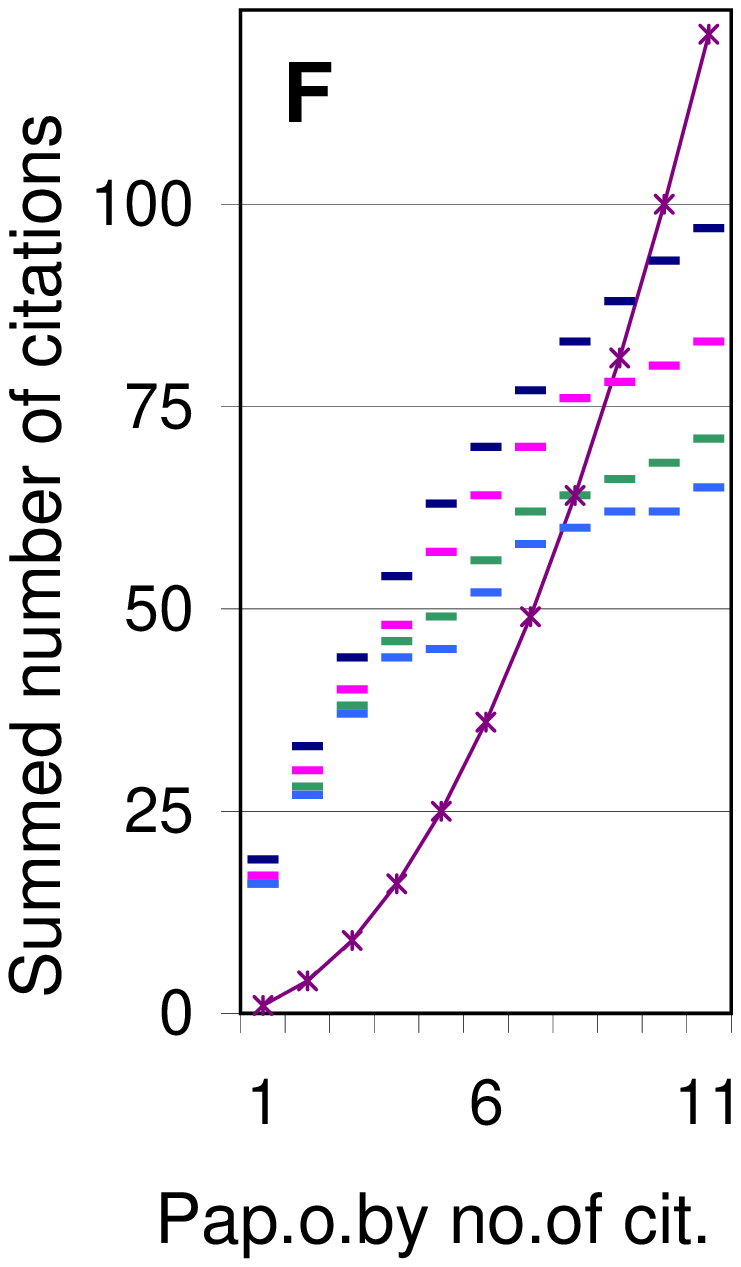}
\\
\vspace{0.5cm}
\includegraphics[height=5.2cm]{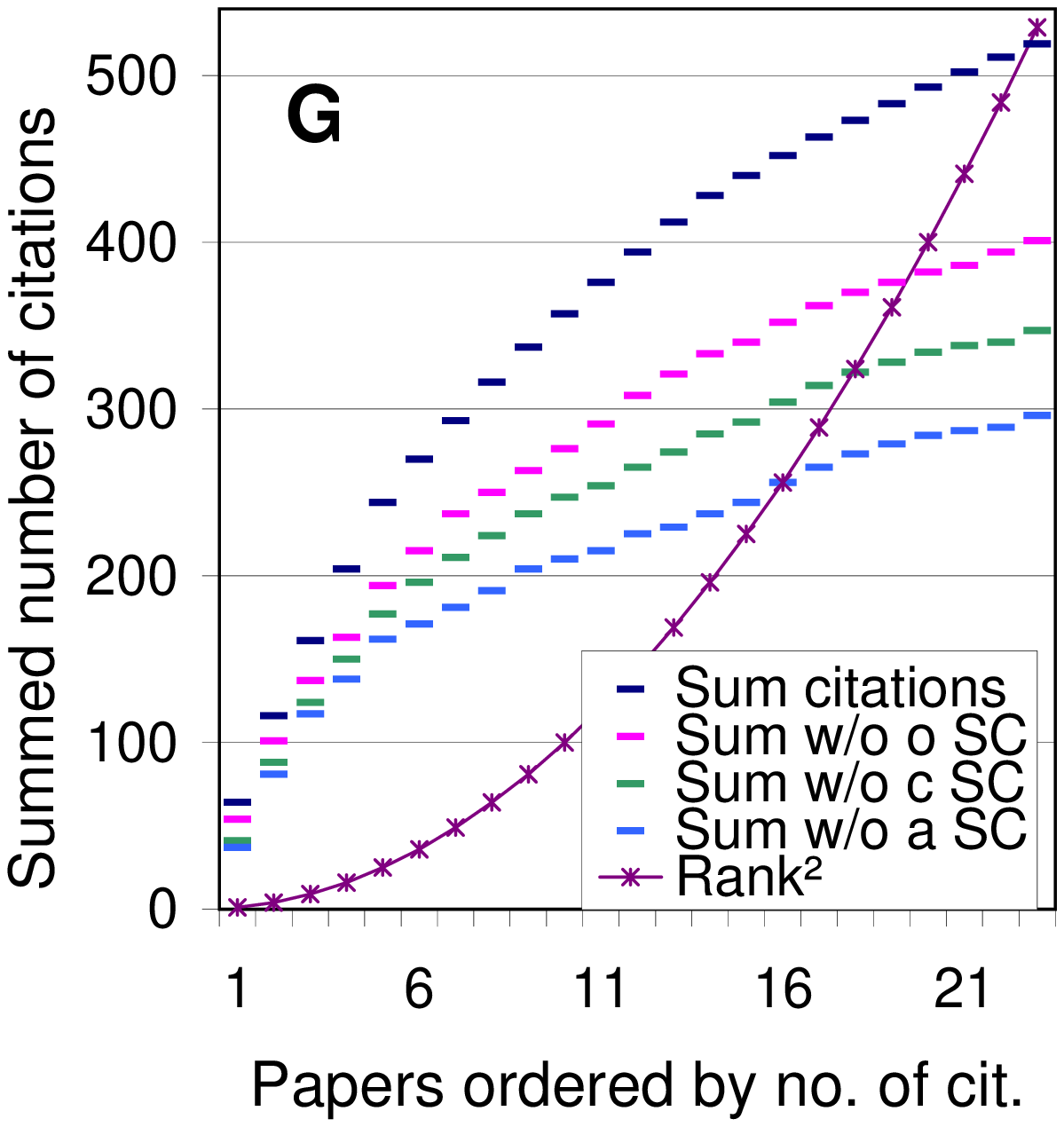}\hspace{2em}\includegraphics[height=5.2cm]{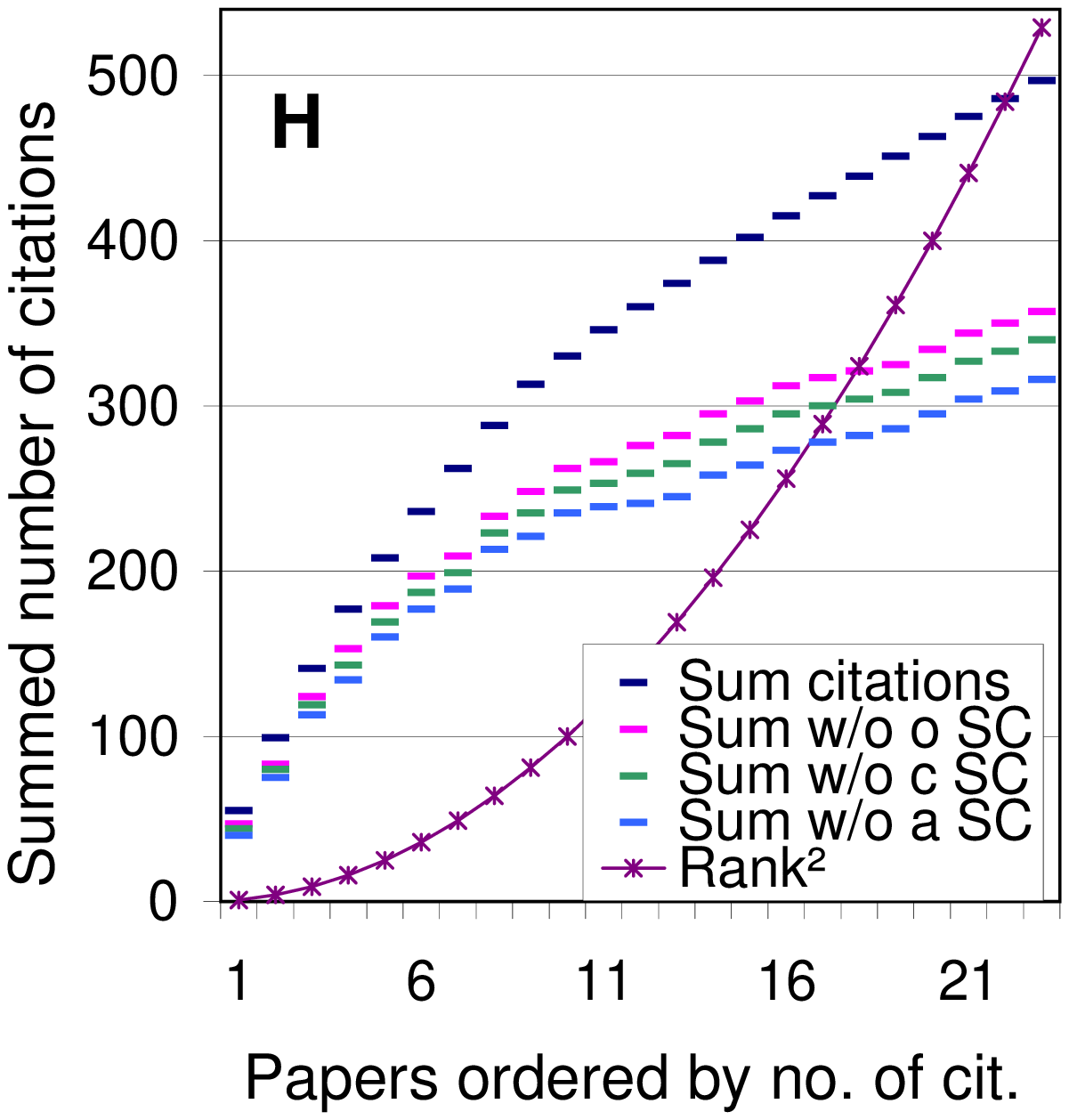}\hspace{2em}\includegraphics[height=5.2cm]{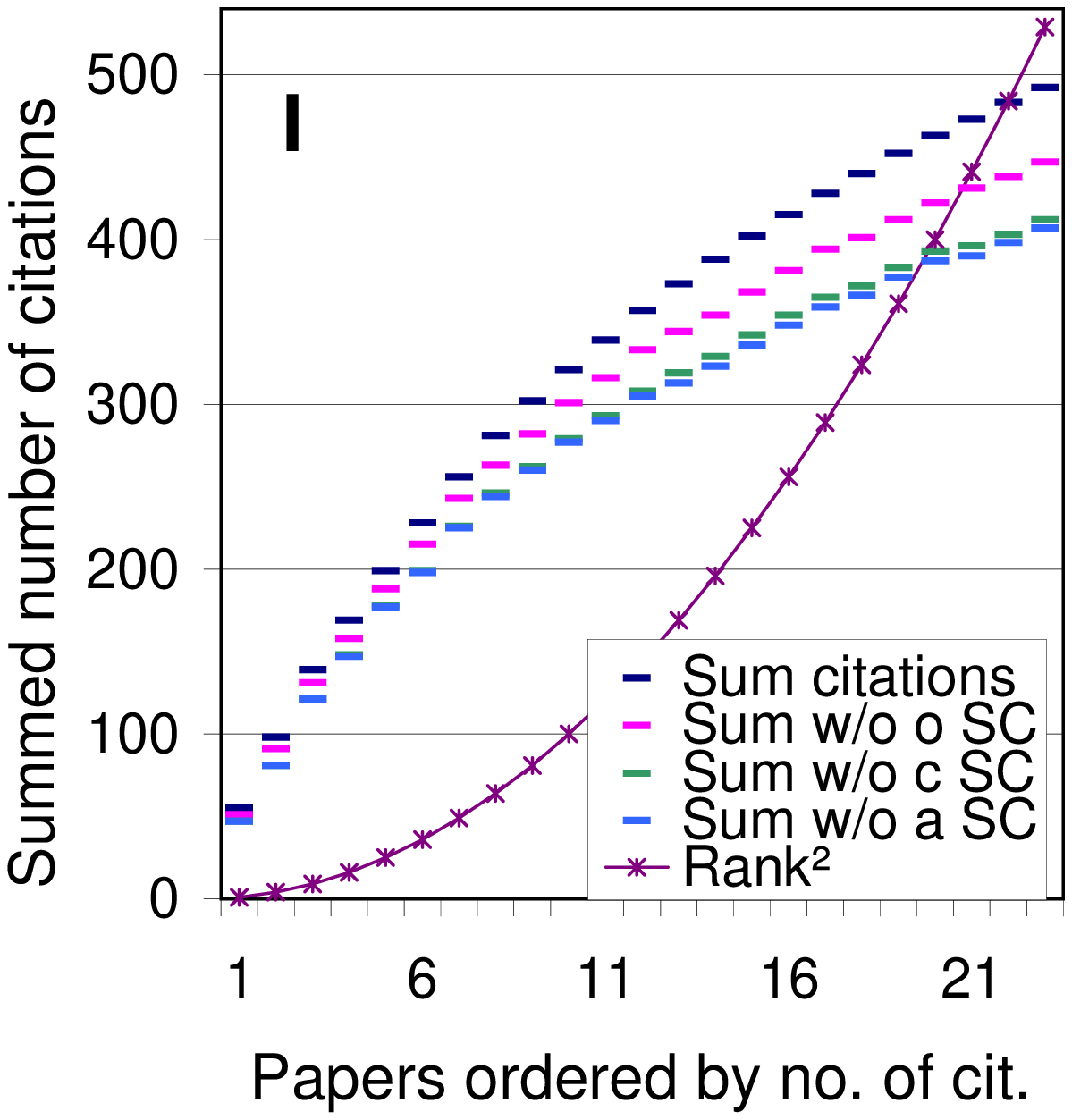}
\caption{Summed number $s(r)$ of citations for the $r$ most cited
papers in each data set (black), without own self-citations (light
grey), without maximal number of any co-author's self-citations
(medium grey), and without all self-citations (dark grey), from
top to bottom. The parabola $r^2$ is also shown. As the SCCs of
the second kind are often equal or not much larger than the SCCs
of the first kind, the respective bars cannot be distinguished in
the plots in particular for small rank. For data set C they are
equal up to $r=15$. The letter in the upper left corner denotes
the data set label.} \label{fig.1}
\end{figure}
\indent In order to exclude the self-citations I have analyzed the
citation records of every manuscript in the nine data sets A -- I
up to the respective $g$ value and somewhat beyond. First I
identified how often the authors cited their own papers, i.e.
direct self-citations \cite{Gl4}. These I call the self-citation
corrections (SCCs) of the first kind and label the respective
quantities with the index $o$ for own SCCs. The respective data
are shown in figure 1 and lead to a significant reduction of the
citation counts by more than 10 \% in nearly all cases, with data
set D being a notable exception for which the own self-citations
amount to 6 \% only; the other exception is data set I with 9 \%
direct self-citations. Although the citation counts of some papers
in data set B suffer from more than 100 self-citations, the
remaining citations by other authors are numerous enough that the
effect on the total number of citations is not overly significant.
On the other hand, for papers 1 and 8 in data set C only three and
one other citations remain after the SCCs of the first kind. In
this case most of the other articles also suffer heavily from
these SCCs. This makes a drastic impact on the respective diagram:
the gap between the full citation count and the reduced count with
the SCCs of the first kind opens immediately, and it is very large
already at the beginning. In my own case (A) relatively few
self-citations of the first manuscripts lead to an unusually small
discrepancy between the full citation count and the one without
SCCs in the beginning. In this aspect, data sets B and E appear to
be more common as confirmed by the cases G, H, and I. As already
mentioned, a notable exception is data set D which shows
relatively few self-citations for nearly all articles so that the
citation count after excluding the self-citations of the first
kind remains relatively close to the total citation count all over
the diagram, and case I shows a similar behavior though less
strongly.
\\
\indent Of course, if a paper is cited by one of the co-authors,
such a citation should also not be taken into account. Considering
the co-author with the highest citation count, which might be the
author him/herself, I have determined these SCCs which I call SCCs
of the second kind and I label the respective quantities with the
index $c$ for co-author SCCs. These data are also shown in figure
1 and for many publications these SCCs of the second kind are not
larger than the SCCs of the first kind. As a result, the
respective reduced citation counts for the first papers in each
diagram of figure 1 are often identical. In the case C this is
true up to $r=15$. A notable exception is the third paper in data
set E where the co-author self-citations are unusually significant
and determine the gap between SCCs of the first and second kind in
the rest of the diagram. For most papers in data set D the SCCs of
the second kind are much larger than those of the first kind,
especially for the first paper, but altogether they remain in the
order of 10 \%, i.e. still relatively small.
\\ \indent
To determine the self-citations of all co-authors, one cannot
simply sum the self-citations of all co-authors, because when two
authors of an article have written another paper together, citing
the first one, this would count as a self-citation for both
co-authors. Rather one has to check every citing paper for
multiple co-author self-citations. I call these the SCCs of the
third kind and label the respective quantities with the index $s$
to denote the correlated sum of all self-citations which yields
the sharpened $g$ index $g_s$. The data are also included in
figure 1. Now all data sets show a significant further reduction
of the citation counts, again with the exception of cases D and
I.\\
\indent Overall, the share of self-citations is of the order of 25
 \% in agreement with previous investigations \cite{Aks,Gl4}. But
in the extreme cases, I found as little as 11.8 \% self-citations
for the first $g$ papers in data set D, and as much as 67.2 \%
self-citations in the $g$-defining set of papers for author C.

\section{Results of the second analysis: the sharpened index $g_s$}
In all diagrams of figure 1 the parabola $r^2$ is also shown. In
principle it allows the determination of the $g$ index from the
intersection of the parabola with the various data curves. To be
precise, the respective $g$ values are given by the rank for which
the parabola is just below or at the plotted bar which reflects
the summed number of citations for that rank. In this way one
finds, e.g., $g_o^A=40$, $g_c^A=39$, and $g_s^A=38$, while for
data set C, one would get $g_o^C=g_c^C=7$ and $g_s^C=5$ or
$g_o^I=g_c^I=17$ and $g_s^I=16$ in the case I. However, these are
preliminary estimates which are not correct because the papers in
figure 1 are ordered by the total number of citations including
self-citations. To obtain the correct values for the $g$ index
with SCCs one has to reorder the papers according to the citation
counts after evaluating the SCCs. Of course, this reordering
should not be restricted to the papers in the $g$-defining set.
This is the reason, why it was necessary to analyze the citation
records ``somewhat beyond`` the rank determined by the $g$ value.
\\ \indent The
reordered data are displayed in figure 2. Here I have chosen a
different way of presentation, the parabola has been already
subtracted from the data, so that now the roots of the resulting
curves $s(r)-r^2=0$ indicate the respective $g$ indices. Of
course, as we are discussing the discrete case, the index is given
by the highest rank for which there is still a non-negative value
of $s(r)-r^2$.

\begin{figure}
\includegraphics[height=5.8cm]{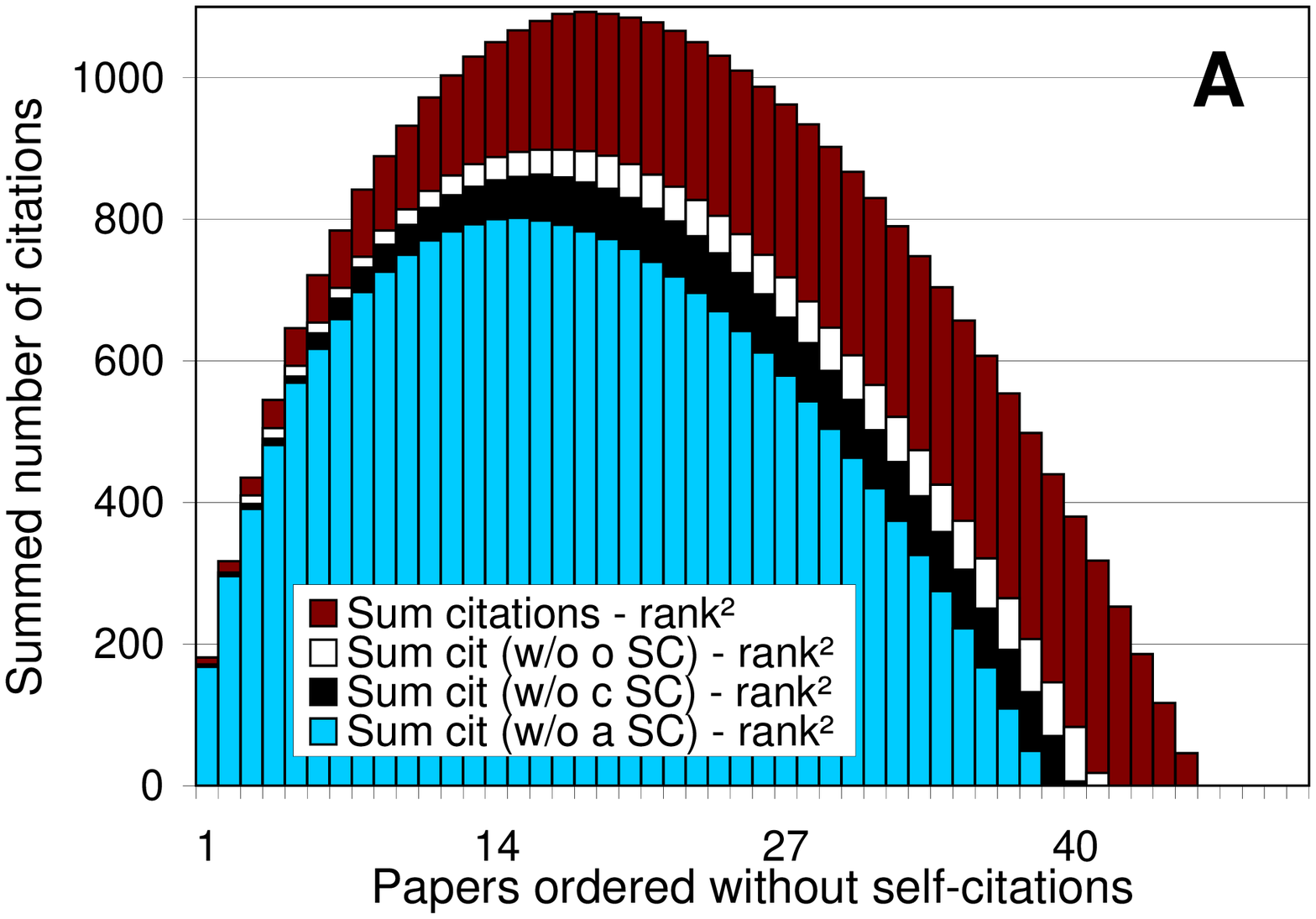}\hspace{2em}\includegraphics[height=5.8cm]{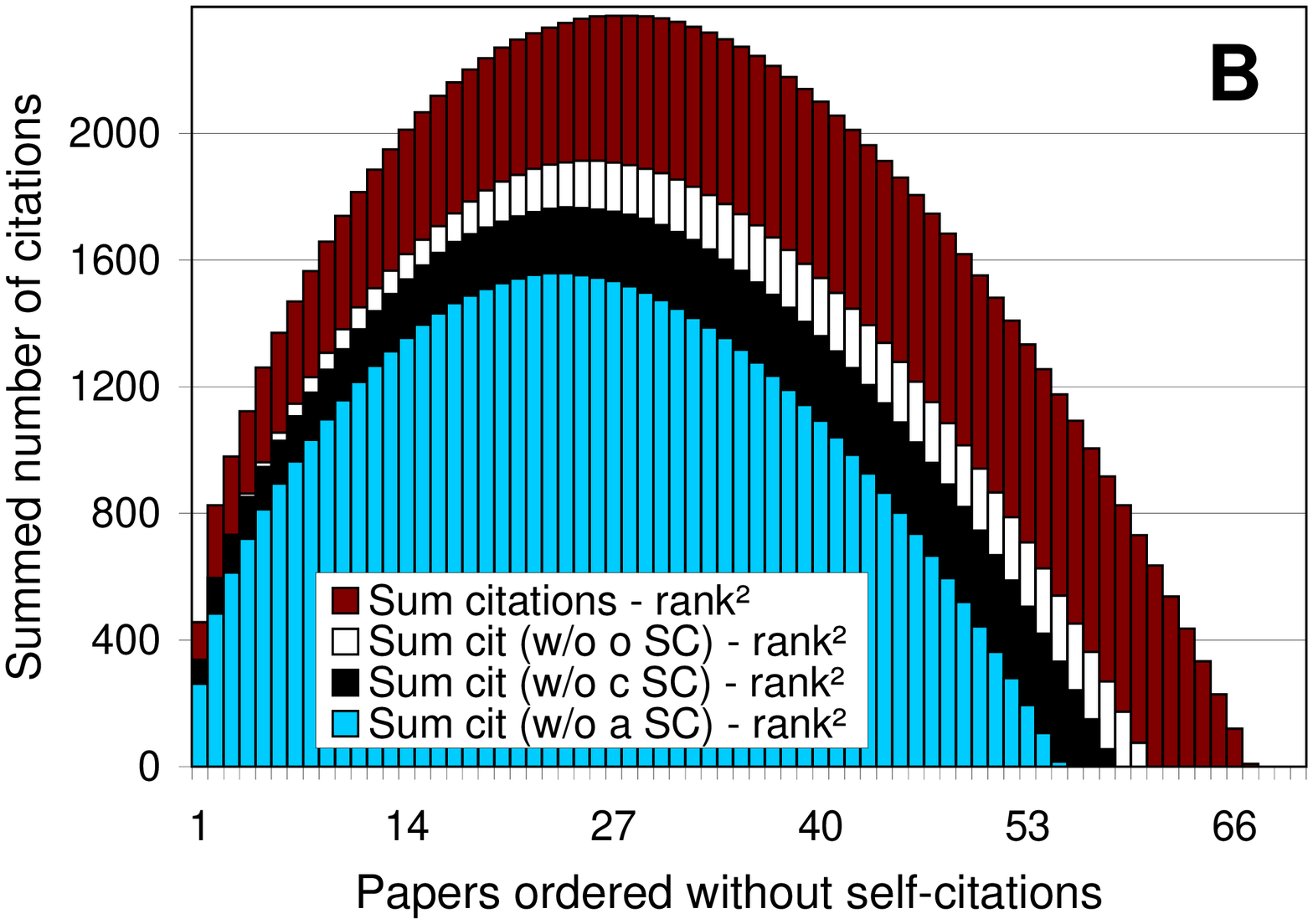}
\\
\vspace{0.5cm}
\includegraphics[height=5.0cm]{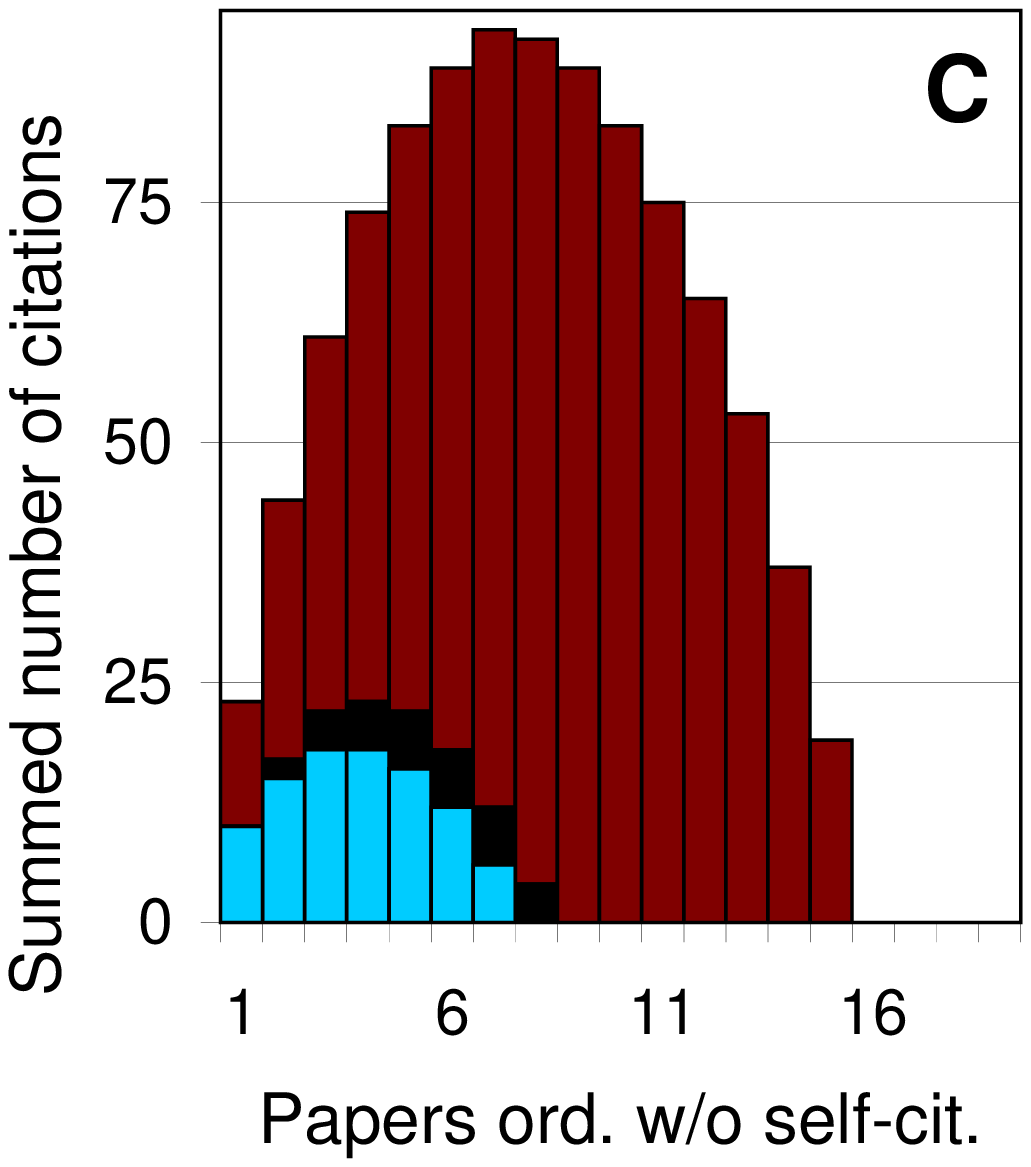}\hspace{2em}\includegraphics[height=5.0cm]{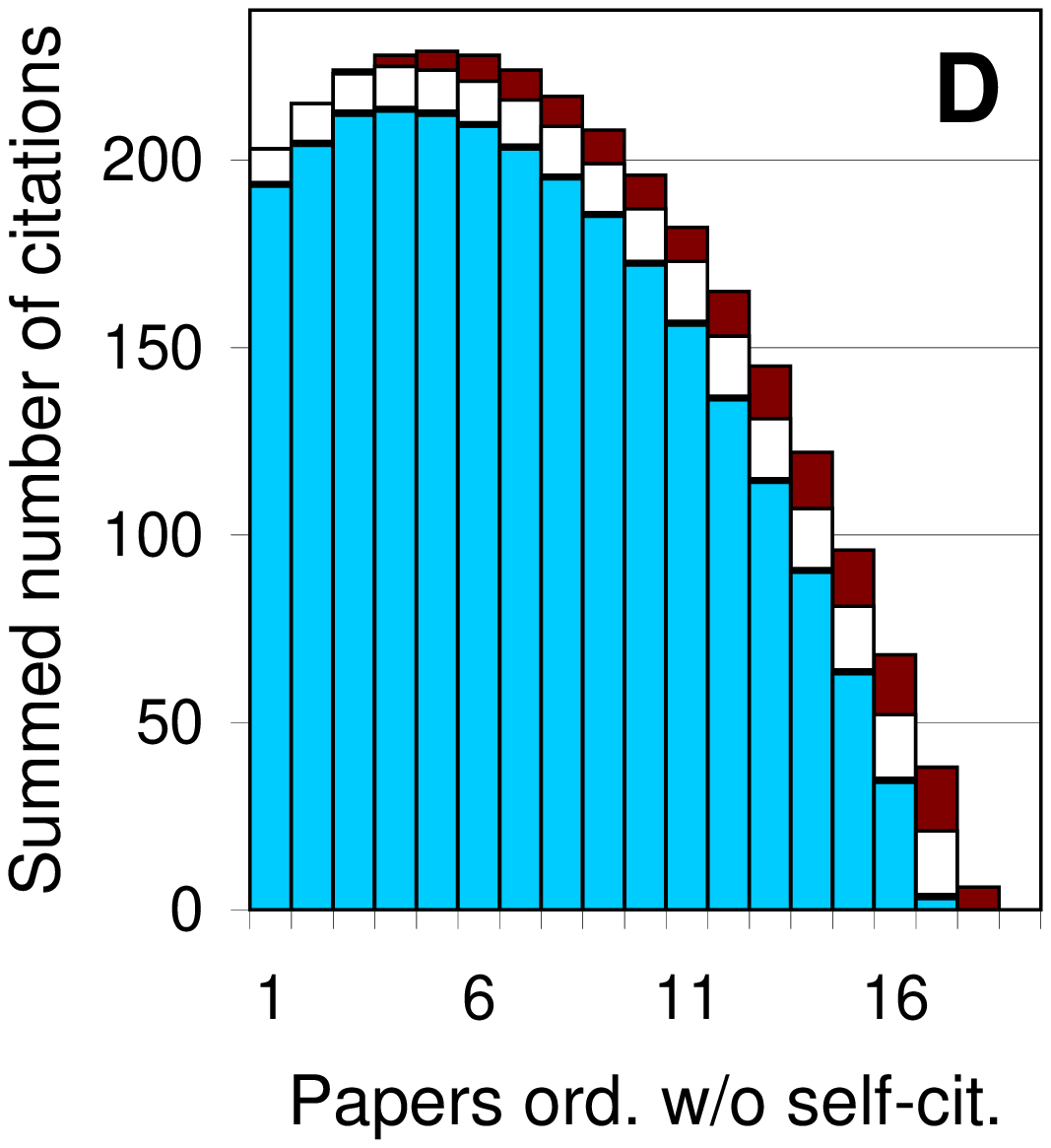}\hspace{2em}\includegraphics[height=5.0cm]{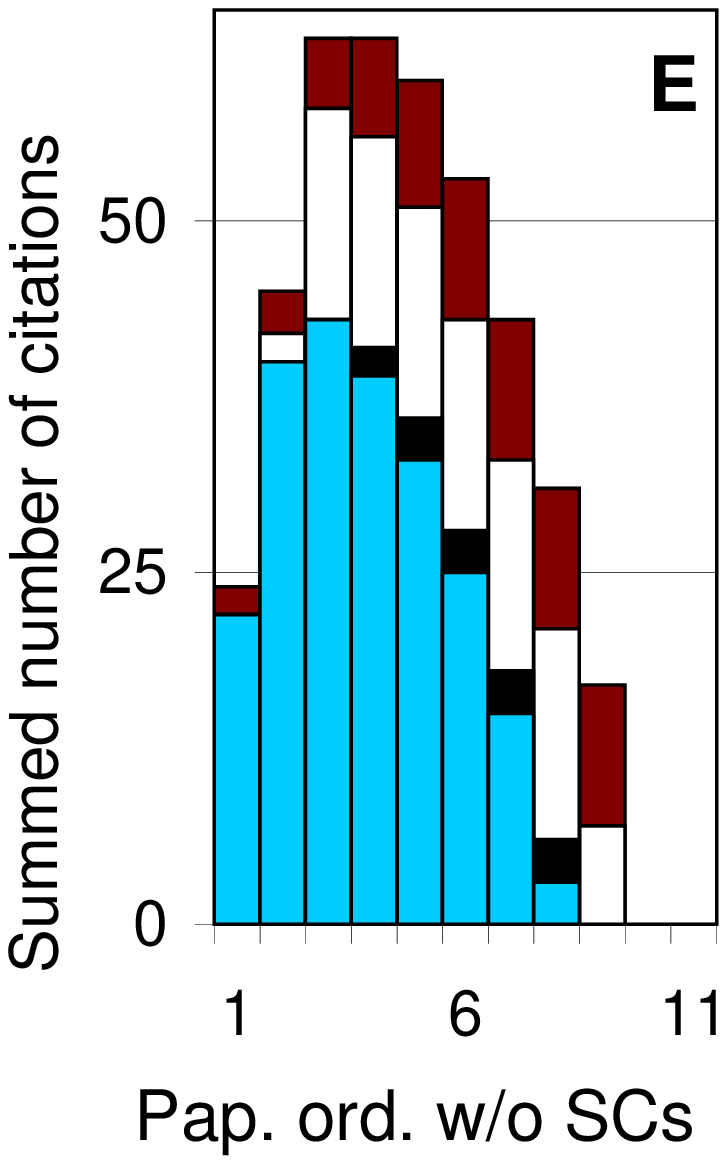}
\hspace{2em}\includegraphics[height=5.0cm]{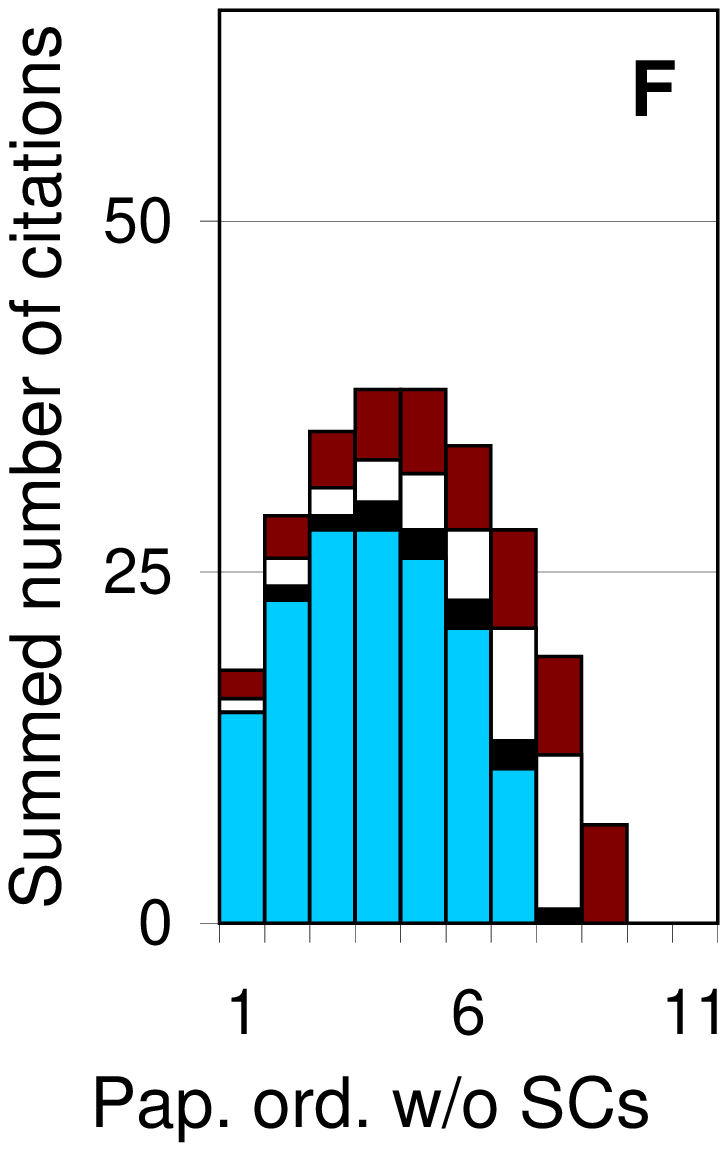}
\\
\vspace{0.5cm}
\includegraphics[height=5.0cm]{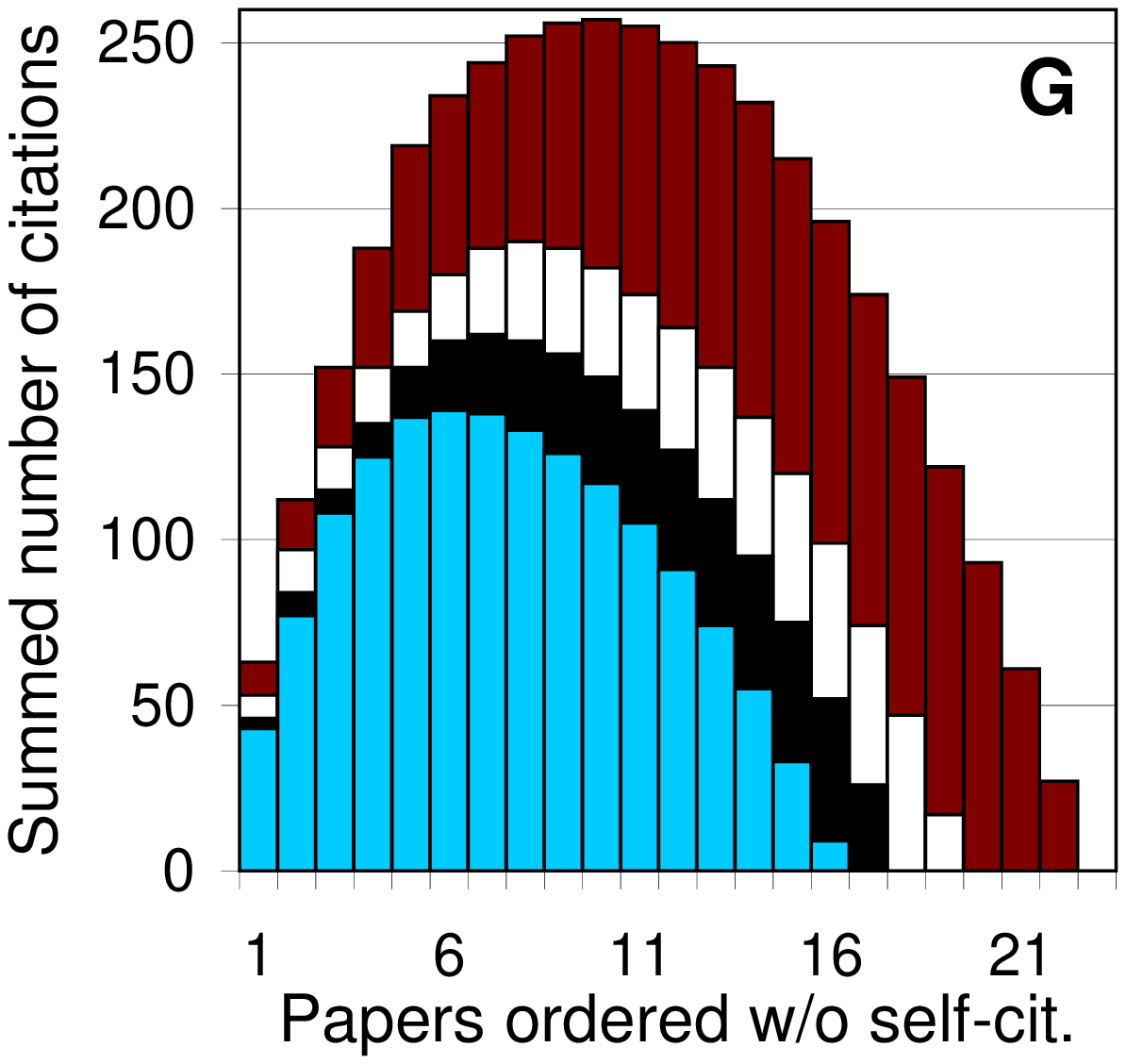}\hspace{2em}\includegraphics[height=5.0cm]{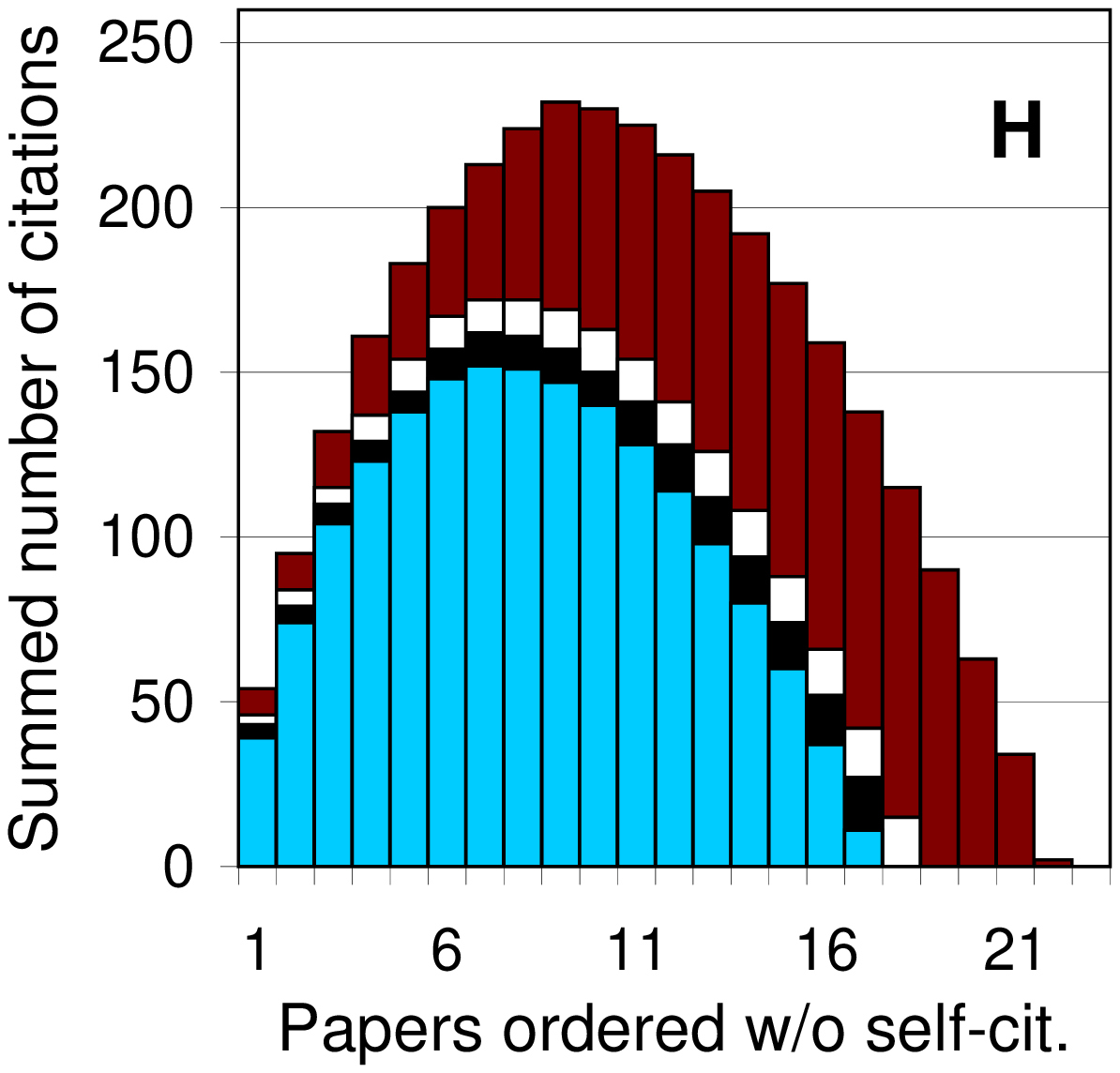}\hspace{2em}\includegraphics[height=5.0cm]{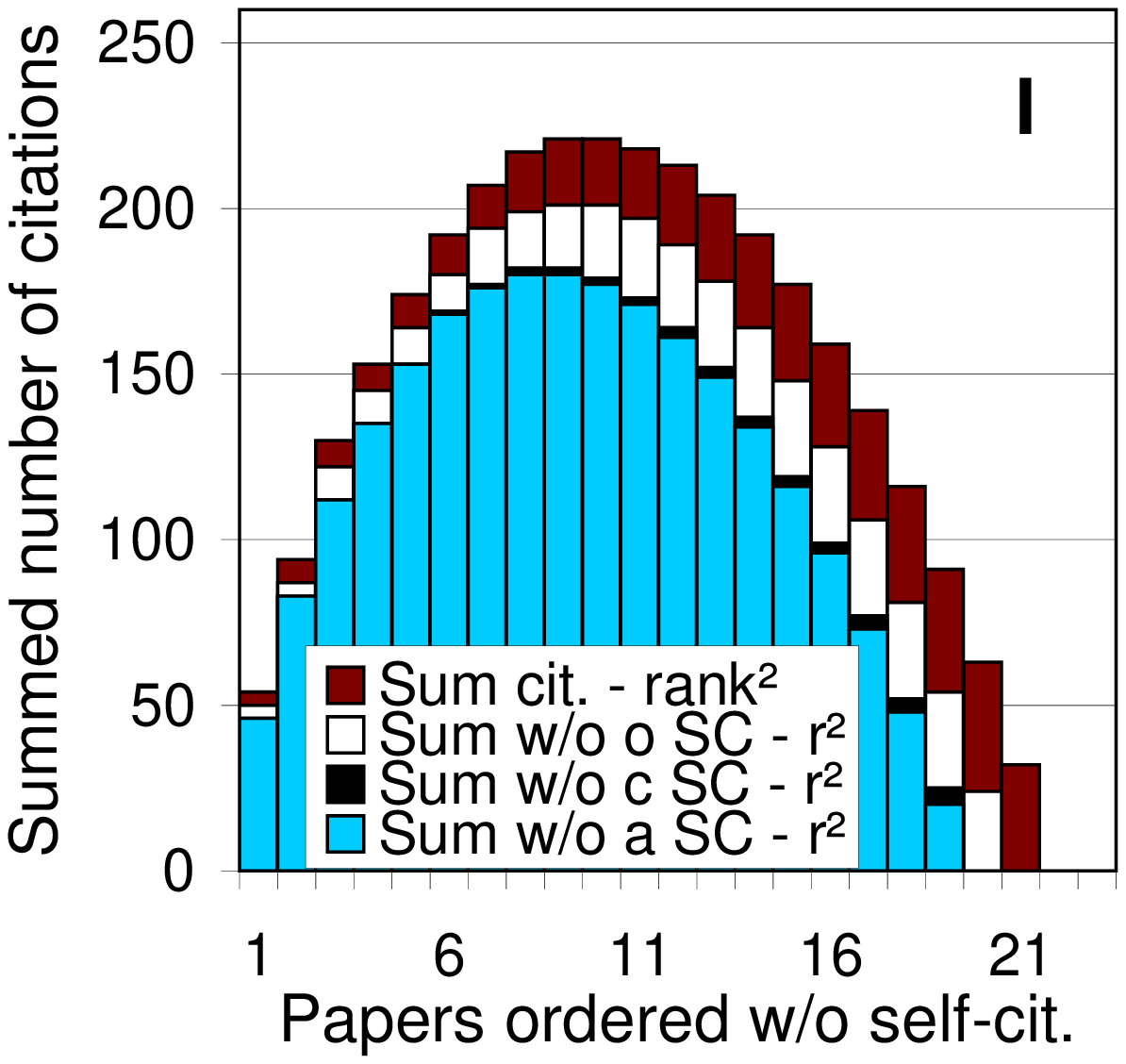}
\caption{Summed number of citations as in figure 1, but reordered
after the exclusion of self-citations and reduced by $r^2$. From
top to bottom: Total number of citations (dark grey), without own
self-citations (white), without maximal number of any co-author's
self-citations (black), and without all self-citations (medium
grey). As the latter histograms conceal the previous ones, the
columns of second and third kind often do not show up, because
they are not different from the third and/or fourth kind. The
letter in the upper right corner denotes the data set label.}
\label{fig.2}
\end{figure}

An advantage of this way of presentation is that the vertical
range of the plots is much smaller so that smaller differences can
be detected more easily. The above observations about the
influence of self-citations can be made in this presentation as
well. Here one can also easily estimate and compare the relative
weight of the various SCCs. Again data set C is an extreme case,
for which even after the reordering very small citation counts
result and where within this range of papers the SCCs of the
second kind are always equal to those of the first kind, so that
no distinction in the respective diagram can be made. In this
presentation case D clearly differs from the other cases, because
the diagram is dominated by the extremely high citation count of
the first paper. A similar distinction can be made for case E, in
which the first 3 papers dominate the total citation count.
Moreover, the unusually large SCCs of the second kind for the
third paper in data set E become most prominent in this
presentation.

\begin{table}[t]
\caption{Influence of self-citations on the Hirsch index $h$ and
on the $g$ index, quantified by the sharpened Hirsch index $h_s$
and the indices $g_o$, $g_c$, and $g_s$ considering the SCCs of
the first, second, and third kind, respectively, as described in
the text, as well as various ratios between these indices.\\}
\label{tab:1}
\renewcommand{\arraystretch}{1.1}
\begin{tabular}{crrrrrrrrrrrrrrr}
  \hline
  \vspace{0cm}
     data set& &$h$&$h_s$&$h_s/h$& &$g$&$g_o$&$g_c$&$g_s$&
     &$g_s/h_s$&$g_s/g$&$g_s/h$&\\
      \hline
 A & & 27 & 22 & 0.81 & & 45 &41 &40 &38 & &1.73 &0.84 &1.41\\
 B & & 39 & 34 & 0.87 & & 67 &60 &58 &55 & &1.62 &0.82 &1.41\\
 C & & 13 & 7 & 0.54 & & 15 & 8 & 8 &7  & &1.00 &0.47 &0.54\\
 D & &  8 & 7 & 0.88 & & 18 &17 &17 &17 & &2.43 &0.94 &2.13\\
 E & &  5 & 3 & 0.60 & & 10 & 9 & 8 & 8 & &2.67 &0.80 &1.60\\
 F & &  7 & 6 & 0.86 & &  9 & 8 & 8 & 7 & &1.17 &0.78 &1.00\\
 G & & 14 & 10& 0.71 & & 22 &19 & 17& 16& &1.60 &0.73 &1.14\\
 H & & 14 &10 & 0.71 & & 22 &18 & 17& 17& &1.70 &0.77 &1.21\\
 I & & 14 &13 & 0.93 & & 21 &20 & 19& 19& &1.46 &0.90 &1.36\\
  \hline
  \end{tabular}
\end{table}

The resulting values for $g_o$, $g_c$, and $g_s$ are given in
table 2. It can be seen that the $g$ index is significantly
reduced in all cases, the sharpened index $g_s$ is mostly reduced
by about 20 \%, but in the extreme cases by as much as 53 \% and
as little as 6 \% only, shown by the ratio $g_s/g$ in table 2. A
comparison of the values in table 2 with the above given
preliminary results for data sets A, C and I as deduced from
figure 1 shows that the reordering does have an influence on the
results. The thus caused small changes of the values for the cases
A and I is typical, the relatively large changes in case C are due
to the unusually large number of self-citations in connection with
the relatively small value of the $g^C$ index.
\\ \indent
Table 2 also comprises the values of the sharpened Hirsch index
$h_s$ and shows the ratio $h_s/h$. With a range of 12 \% to 46 \%
the reduction is of the same order as for the sharpened index
$g_s$. However, the variance is somewhat higher for $g_s$ than for
$h_s$. Therefore also the ratio $g_s/h_s$ between the two
sharpened values shows a larger variance ranging from 100 \% to
267 \% and thus makes a comparison between the authors concerning
their visibility clearer. This becomes even more apparent, when
one relates the sharpened value $g_s$ with the original Hirsch
index: now the ratio $g_s/h$ yields values between 0.54 and 2.13.
The coincidence of $g_s^C$ and $h_s^C$ appears surprising at a
glance, because it seems to indicate the unrealistic exception
$g_s=h_s$. However, this is not the case, rather the equality
stems from the fact that we are considering integer values only so
that in this particular case a total citation count of
$s_s^C(g_s^C)=55$ yields $g_s^C=7$ and not $g_s^C=7.4$.
Nevertheless, the coincidence is an indicator that the citation
counts of all manuscripts in the $h_s^C$-defining set are at or
not much above $h_s^C$. The small number of self-citations in data
set D is already reflected in the high value $h^D_s/h^D$, but it
becomes even clearer from the ratio $g^D_s/g^D$ being close to
unity. Similarly, the other extreme $h_s^C/h^C=0.54$ becomes more
pronounced in the $g$ index ratio $g_s^C/g^C=0.47$.

\section{Comparison of different data sets}
Let me finally point out various observations that can be made by
comparing some of the values for different data sets. For example,
my colleague B has written a few more papers than I did, but as he
is working in a very topical field, his publications have received
significantly more citations as reflected not only in the citation
count of the mostly cited paper, but also in the summed number of
citations. Consequently his Hirsch index is significantly larger
than mine. All other indices, in particular $g$, $h_s$, and $g_s$,
show a very similar behavior for both cases and all the ratios in
table 1 and table 2 are quite similar.\\ \indent Comparing data
sets C and D the total number of citations and the Hirsch index
give the impression that my colleague C is much more productive
than D. However, the sharpened Hirsch index is equal for both
cases, the $g$ index is already smaller for the data set C and the
sharpened $g$ index makes this distinction even more obvious. One
may conclude that D is more visible than C, although C is more
productive than D.\\ \indent A similar picture can be painted
comparing cases E and C. The total number of publications in data
set E is very small, so is the Hirsch index, although this
scientist has already obtained an associate professorship.
Certainly the value $h^E=5$ is much smaller than the value which
Hirsch has proposed \cite{Hirsch} as a reasonable value for
promotion. It is also much smaller than the already discussed
value $h^C$. However, table 1 shows that the highest citation
counts of these two cases are nearly the same. This is also true
for their $a$ indices. Comparing the $g$ values, one finds that
$g^E$ approaches $g^C$, which is also reflected in the respective
ratios $g/h$. The ratios $h_s/h$ are not so much different, but
the ratios $g_s/g$ are clearly distinct again, reflecting the
observation that in the sharpened $g$ index the colleague C has
been overtaken by E. In the ratio $g_s/h_s$ this becomes quite
obvious, and it is even more pronounced when one compares the
ratios $g_s/h$. Again the conclusion is that the impact of the
research of scientist E is at least comparable, if not larger than
that of colleague C. However, due to the small numbers of the
indices in these two cases, the relative changes and the relative
differences should not be overestimated.\\
\indent Comparing data sets F and E, a similar observation can be
made: while colleague F appears to be more productive than E with
more publications and more visible with a higher $h$ index, the
relation is reversed when one looks at the $g$ index or at the
sharpened index $g_s$. Comparing cases F and C, F appears less
productive and less visible, in this case not only by comparing
the $h$ values but also the $g$ values, but this difference is
completely balanced in the sharpened values of $g_s$.\\
\indent The three cases G, H, and I have been selected, because
they have approximately the same total number $n$ of publications
and similar highest citation counts $c(1)$. It was therefore not
surprising, that their $h$ indices are equal and that their $g$
indices are nearly the same. However, after the SCCs have been
taken into account, the resulting $g_s$ values are different and
allow a distinction. The colleague G with most publications ends
up with the smallest value $g_s^G=16$, while colleague I with
smallest number of cited publications and the smallest $g$ index
among these three cases still reaches the value
$g_s^I=19$.\\

\section{Summary}
In conclusion, I have presented a case study of the $g$ index as
introduced by Leo Egghe. The comparison with the respective values
for the Hirsch index confirmed the expectation that the $g$ index
does indeed incorporate the evolution of the citation counts of
highly cited articles, which are not appropriately appreciated in
the Hirsch index. In principle the determination of the $g$ index
is as simple as that of the $h$ index, both being single numbers.
In practice it is somewhat more difficult to determine, because it
is an integral measure.\\ \indent In addition, the influence of
self-citations on the $g$ index has been analyzed. As already
observed for the Hirsch index \cite{MS} the respective corrections
are not insignificant and can drastically reduce the $g$ index.
Sharpening the $g$ index by the exclusion of the self-citations
leads to a larger variance of the obtained values and thus allows
a clearer distinction of the data sets. \\ \indent The expenditure
for the data analysis is much larger, when the self-citations are
to be determined. This was mentioned as one reason, why the
self-citations had been taken into account only in seven randomly
selected cases out of 187 data sets of scientists of ecology and
evolution, for which the Hirsch index was investigated \cite{Kel}.
This expenditure is even much larger for the determination of the
$g_s$ index than for the calculation of the $h_s$ index, because
now the citation counts of all manuscripts at least up to the
critical rank $g$ have to be analyzed and not only the
self-citations for the manuscripts with a rank in the neighborhood
of the Hirsch index $h$. On the other hand, while small errors can
easily change the Hirsch index $h$, this is usually not the case
for the $g$ index, because of its integral character. This has
become clear, when the manuscripts were reordered by the number of
their citations after excluding the self-citations and when this
reordering did not have a significant influence while such a
reordering is essential for the sharpening of the Hirsch index \cite{MS}.\\
\indent In conclusion, the present study corroborates the
superiority of the $g$ index as compared to the $h$ index. It also
confirms that self-citations can be very significant and should be
excluded, thus sharpening the $g$ index and its merit. \\

\section{Acknowledgement}

I thank A. Clau{\ss}ner and C. Schreiber for tedious work in
establishing the data base.


\begin{thebibliography}{10}

\bibitem{Hirsch}
J.E. HIRSCH, An index to quantify an individual's scientific
research output, {\it Proc. Natl. Acad. Sci. U.S.A.} {102} (2005)
{16569 -- 16572}.

\bibitem{Pop}
S.B. POPOV, {A parameter to quantify dynamics of a researcher's
scientific activity}, {\it arXiv:physics} /0508113.

\bibitem{Leh}
S. LEHMANN, A.D. JACKSON, B. LAUTRUP, {Measures and mismeasures of
scientific quality}, {\it arXiv:physics \rm/0512238 and Measures
for measures, \it Nature}, {444} (2006) {1003 -- 1004}.

\bibitem{Bat}
P.D. BATISTA, M.G. CAMPITELI, O. KINOUCHI, A.S. MARTINEZ, Is it
possible to compare researchers with different scientific
interests?, {\it Scientometrics} {68} (2006) {179 -- 189}.

\bibitem{Eg5}
L. EGGHE, Theory and practise of the $g$-index, {\it
Scientometrics} {69} (2006) {131 -- 152}.

\bibitem{Pod}
I. PODLUBNY, K. KASSAYOVA, Towards a better list of citation
superstars: compiling a multidisciplinary list of highly cited
researchers, {\it Research Evolution} {15} (2006) {154 -- 162}.

\bibitem{Meho}
L. MEHO, The rise and rise of citation analysis, {\it Physics
World} {20, no. 1} (2007) {32 -- 36}.

\bibitem{Bih}
B.H.~JIN, H-index: An evaluation indicator proposed by scientist,
{\it Science Focus} {1} (2006) {8 -- 9}.

\bibitem{Eg2}
L. EGGHE, An improvement of the h-index: the g-index, {\it ISSI
Newsletter} {2} (2006) {8 -- 9}.

\bibitem{Eg4}
L. EGGHE, How to improve the h-index, {\it Scientist} {20} (2006)
{14}.

\bibitem{Ros}
R. ROUSSEAU, {Simple models and the corresponding h- and g-index},
{\it Science Technology Development}, to appear (2007); {\it
http://eprints.rclis.org/archive}/00006153.

\bibitem{Roe}
H.L. ROEDIGER, The h index in science: A new measure of scholarly
contribution, {\it APS Observer} {19, no. 4} (2006).

\bibitem{Aks}
D.W. AKSNES, {A macro-study of self-citation}, {\it
Scientometrics} {\bf 56} (2003), 235 -- 246.

\bibitem{Sny}
H. SNYDER, S. BONZI, {Patterns of self-citation across
disciplines}, {\it Journal of Information Science}, {\bf 24}
(1998), 431 -- 435.

\bibitem{Gl4}
W. GL\"ANZEL, B. THIJS, B. SCHLEMMER, {A bibliometric approach to
the role of author self-citations in scientific communication},
{\it Scientometrics} {\bf 59} (2004), 63 -- 77.

\bibitem{MS}
M. SCHREIBER, Self-citation corrections for the Hirsch index, {\it
EPL} {78} (2007) 30002: 1 -- 6.

\bibitem{Kel}
C.D. KELLY, M.D. JENNIONS, The h index and career assessment by
numbers, {\it Trends in Ecology and Evolution} {21} (2006) {167 --
170}.

\bibitem{Cro}
B. CRONIN, L. MEHO, Using the h-index to rank influential
information scientists, {\it J. Am. Soc. Inf. Sci. Techn.} {57}
(2006) {1275 -- 1278}.


\end{thebibliography}
\end{document}